\documentclass[10pt,a4paper,english,nofootinbib,twocolumn]{revtex4}
\usepackage{lmodern}

\usepackage[T1]{fontenc}
\usepackage[latin9]{inputenc}
\usepackage{babel}
\usepackage{amsmath}
\usepackage{amssymb}
\usepackage{esint}
\usepackage{graphicx}
\usepackage[unicode=true,pdfusetitle,
 bookmarks=true,bookmarksnumbered=false,bookmarksopen=false,
 breaklinks=false,pdfborder={0 0 1},backref=false,colorlinks=false]
 {hyperref}

\makeatletter

\@ifundefined{textcolor}{}
{%
 \definecolor{BLACK}{gray}{0}
 \definecolor{WHITE}{gray}{1}
 \definecolor{RED}{rgb}{1,0,0}
 \definecolor{GREEN}{rgb}{0,1,0}
 \definecolor{BLUE}{rgb}{0,0,1}
 \definecolor{CYAN}{cmyk}{1,0,0,0}
 \definecolor{MAGENTA}{cmyk}{0,1,0,0}
 \definecolor{YELLOW}{cmyk}{0,0,1,0}
}

\@ifundefined{definecolor}{\usepackage{color}}{}
\usepackage{babel}

\usepackage{latexsym}\usepackage{bm}

\makeatother

\usepackage{babel}
\begin{document}

\title{Inflation in pure gravity with only massless spin-2 fields  }

\author{ Bayram Tekin}

\email{btekin@metu.edu.tr}

\selectlanguage{english}%

\affiliation{Department of Physics,\\
 Middle East Technical University, 06800 Ankara, Turkey}

\date{\today}
\begin{abstract}
We show that without introducing additional fields or extra degrees of freedom, a specific higher derivative extension of Einstein's gravity that has only a massless spin-2 excitation in its perturbative  spectrum, has an inflationary period, a quasi-de Sitter phase with  enough number of $e$-foldings required to solve the horizon and related problems. The crucial ingredient in the construction is the curvature dependence of the effective Newton's constant.

\end{abstract}
\maketitle

\section{Introduction}

The Cosmic Microwave Background (CMB) data shows an extreme homogeneity in the apparently causally disconnected 
regions in the sky \cite{wmap}. In some sense, according to the CMB picture, the Universe looks a lot older than it really is. In principle, there are two main lines of reasoning to explain this: either  our counting and notion of the "time"  at the birth of the time and the birth of universe is completely wrong and the correct picture of time that will emerge in full " quantum gravity" theory is so dramatically different that the Universe really had enough time to solve the horizon and related problems. This point of view outcasts the problems to the putative quantum gravity, while it may be a viable solution, there is not much one can do about it right now in the absence of a proper theory. Or, as a second, more conventional, option the universe went through a brief inflationary period (an accelerated expansion phase with a shrinking comoving Hubble radius) which can be comfortably handled within a classical space-time picture governed by general relativity with a caveat \cite{Abbott}. The inflationary expansion should be brief, meaning pure general relativity with a positive cosmological constant is ruled out as it predicts inflation forever and hence the caveat : with conventional matter (renormalizable fields in the standard model) one does not  have an inflating universe with correct amount of friction to stop inflation.  To solve the problem one adds additional fields, such as a scalar field, "the inflaton". But without the constraints coming from the stringent requirements of renormalization in quantum field theory,  the possibilities  of self-interaction (the potential ) of the scalar field is practically endless. So with  additional fields one can have a successful  inflationary period as well as required tiny inhomogeneities, but without the lack of a proper fundamental principle, there is no way to find what these fields are. Hence the plethora of the theories exhibiting  inflationary epochs in the early universe. 

Another route to inflation is to modify gravity at short distances. One of the earliest and most successful  approaches is that of Starobinsky \cite{Starobinsky}  adopting a quadratic modification of  general relativity Lagrangian as  ${\cal{L}} =  R  + b R^2 $ which yields consistent observables today, such as the scalar tilt and scalar/tensor ratio of the fluctuations \cite{Ade}. 
But, overall, this model is equivalent to- after a conformal transformation of the metric-general relativity coupled to a scalar field with a specific potential. This is expected since, the quadratic theory not only has a massless spin-2 graviton, but also a massive spin-0 graviton in its spectrum, and the latter becomes a massive scalar field in the conformally related  metric frame. Many successful  inflation models seem to be related to the Starobinsky model \cite{Alex}, hence to general relativity with a scalar field of some specified (exponential) self-interaction which has no place in a renormalizable four dimensional quantum field theory. 

In this work we propose and pursue a new approach,  built on the wisdom  of conventional inflationary scenarios that the spacetime during inflation is a classical pseudo-Rimennian manifold described by the metric, but is distinctly different from the earlier approaches in one important aspect: we propose that no new fields (such as an inflaton) and no additional degrees of freedom (such as the ones in quadratic gravity) are to be introduced to Einstein's gravity. This line of reasoning can be considered as a bottom-up approach of model building and can be briefly explained as follows.

While we do not know what the full (microscopic) quantum gravity theory
is, we can still ask the following question: What properties should
a \emph{low energy} quantum gravity theory have, as long as the curved
space formulation of gravity is correct; namely, as long as gravity
is described by a metric and spacetime is effectively a pseudo-Riemannian
manifold? A possible answer to this question could come from the lowest
order theory, that is Einstein's general relativity (GR). GR is obviously diffeomorphism invariant,
a property which is very easy to implement in other low energy quantum
gravity theories--cousins of GR that work better at higher energies--but
not a sufficiently restrictive one. There are two other properties
of GR which are usually not discussed in building low energy quantum
gravity theories: 1- The \emph{uniqueness} of the maximally symmetric
vacuum which is the flat space in the absence of a cosmological constant
and the (anti)-de Sitter space in the presence of it. 2- The \emph{masslessness}
of the unitary graviton as an excitation about the vacuum. According
to the accepted wisdom, at higher energies the Lagrangian density
of GR will be augmented with more powers of curvature which immediately  
ruin the above mentioned two properties. For example, in general when $R^n$ terms 
are added, generically one has $n$ maximally symmetric solutions (for $n>2$) and depending on the details of the added terms, 
many new excitations about any one of the vacua. It is very hard to make these additional excitations unitary, namely, non-ghosts and non-tachyons. For example, the quadratic theory  ${\cal{L}} =  R  + b R^2 + a R_{\mu \nu} R^{\mu \nu}$ is renormalizable but has a massive spin-2 ghost and a massive spin-0 particle besides the massless spin-2 graviton \cite{Stelle}. Recently \cite{Gullu4d} , a higher derivative  theory with arbitrary powers of curvature that share the two salient features of general relativity was found. Here, we show that, this minimal extension of general relativity, without the need of additional fields, has an inflationary period, a quasi-de Sitter phase with  enough number of $e$-foldings required to solve the horizon and related problems. Let us briefly introduce the theory.

\section{The Spin-2 Theory: Minimal Extension of GR}

Our rather long search for a theory that has a unique maximally symmetric solution and a single massless spin-2 excitation about this vacuum, as a minimal extension of Einstein's gravity, yielded the following Lagrangian density   in four dimensions \cite{Gullu4d}
\begin{equation}
{\cal L}=\frac{1}{16 \pi G_0\gamma}\left[ \sqrt {\det\left(I+4\gamma A \right)}-\left(4 \gamma\Lambda_{0}+1\right)\right].\label{eq:Generic_BI_intro}
\end{equation}
Note that the action is diffeomorphism invariant and given as  $I = \int d^4 x \sqrt{-\det{g_{\mu \nu}}}\, {\cal L}$. Here,  $A$ is the minimal symmetric  matrix $(A^\mu_\nu)$  given as  
\begin{eqnarray}
&A_{\mu\nu}= R_{\mu\nu} + c S_{\mu\nu} \\
& + 4 \gamma \left( a C_{\mu\rho\nu\sigma} R^{\rho\sigma} + \frac{c + 1}{4} R_{\mu\rho} R_{\nu}^{\rho} + \left( \frac{c \left(c+2\right)}{2} - 2 - b \right) S_{\mu\rho} S_{\nu}^{\rho} \right) \nonumber \\
& +\gamma g_{\mu\nu} \left( \frac{9}{8} C_{\rho\sigma\lambda\gamma} C^{\rho\sigma\lambda\gamma} - \frac{c}{4} R_{\rho\sigma} R^{\rho\sigma} + b S_{\rho\sigma} S^{\rho\sigma} \right), \nonumber 
\label{EGB_Amn} 
\end{eqnarray} 
where $a$, $b$ and $c$ are arbitrary real numbers and $\gamma$ is the Born-Infeld parameter with dimensions of $[L^2]$.  
Its numerical value will be fixed approximately  to the scale the inflation takes place which is several orders of magnitude smaller than the Planck scale. Here $C_{\mu \nu \sigma \rho}$ is the Weyl-tensor and $S_{\mu \nu}$ is the traceless Ricci tensor. 
The properties, the construction of this theory and its $n$ dimensional generalization are noted elsewhere \cite{tahsin3}  based on \cite{tahsin1}, hence we just recapitulate some of the results below.  Defining the dimensionless cosmological constant as  $\lambda \equiv \gamma \Lambda$ and the bare one as $\lambda_0 = \gamma \Lambda_0$,  for any values of $a,b,c$ and $\gamma >0$,
\begin{itemize}

\item The theory has only a single massless spin-2 particle in spectrum about its flat or (anti)-de Sitter space.

\item Effective (dimensionless) cosmological constant of the maximally symmetric solution ($R^{\mu \nu}\,{\sigma \rho} = \frac{\lambda}{3 \gamma} ( \delta^\mu_\sigma \delta^\nu_\rho- \delta^\mu_\rho \delta^\nu_\sigma) $ obeys the equation  
\begin{equation}
4 \lambda^4 + 4 \lambda^3 - \lambda + \lambda_0=0.
\label{eff_cosmo}
\end{equation}

\item  Effective Newton's constant is related to the bare one and $\lambda$ via 
\begin{equation}
 \frac{1}{{G_N}}=\frac{1}{{G_0}}\left(1-4 \lambda\right)\left(1+2\lambda\right)^{2}.
\label{eff_newton}
\end{equation}

\item The theory when expanded in  small curvature  ($\vert \gamma R \vert \ll 1$), first reproduces GR as demanded by low energy observations, and at the quadratic order, it reproduces  GR modified with the Gauss-Bonnet theory, which at the classical level is just GR and at the cubic and the quartic order  specific theories studied recently \cite{esin}.
\end{itemize}

Several pertinent issues need to be mentioned here: the line of thought that led to this theory started with long time ago with Eddington's proposal \cite{Eddington} of writing determinantal actions in the Palatini formulation which have picked up momentum 
both in the metric \cite{Gibbons} and Palatini form in the recent works \cite{Banados,Delsate,Fiorini,Lavinia}. In the metric formulation, the major stumbling block was to construct unitary theories about the (anti)-de Sitter vacua, since the flat case is not restrictive enough. In what follows we shall study the quasi-de Sitter solution in this theory very early in the universe where higher curvature terms cannot be neglected. 
\section{Inflationary Solution}

Let us consider the simple spatially flat Robertson-Walker solution as favored by the data
\begin{equation}
ds^2 = - dt^2 + a(t)^2 d\vec{x} \cdot   d\vec{x}.
\label{RW}
\end{equation}
the field equations coming from the variation of (\ref{eq:Generic_BI_intro}) are complicated \cite{Gullu4d} and are not needed for our current purposes. The usual procedure is to insert the metric into the action and define the Hubble parameter as $H(t) = \frac{\dot{a}}{a}$ and then derive a single equation containing the Hubble parameter and its time derivatives. But this procedure of reducing the action requires extreme care as the symmetry group "integrated-out" here is not a compact one  and so not covered by Palais' symmetric criticality  guarantee \cite{Palais,short}. One cannot simply vary the action with respect to the scale factor or the Hubble parameter as these will yield necessary but not sufficient conditions. [Even the recent literature is full of this subtle mistake which leads to metrics that do not solve the $00$ component of the field equations.]. Weinberg's computation \cite{Weinberg} is robust and error-free which we follow.  Inserting {\ref{RW}} to (\ref{eq:Generic_BI_intro}) and dropping an overall  irrelevant constant we arrive at
\begin{equation}
I =  \int dt \, a^3\, {\cal I}( H, \dot H),
\end{equation}
where we skipped the arguments of the functions which all depend on the coordinate time $t$. The remarkable property of  our theory is that just like Einstein's gravity the second time derivative of the Hubble parameter does not appear in the reduced action in contrast to other higher derivative theories.  One finds that 
\begin{eqnarray}
{\cal I}( H, \dot H)= -1 - 4 \lambda_0 + \sqrt{ X_1^3 X_2}
\end{eqnarray}
where, after defining a dimensionless Hubble parameter  $h(t) \equiv \sqrt{\gamma} H(t)$,  and a dimensionless time $\tilde{t} \equiv  t/\sqrt{\gamma}$ one has 
\begin{equation}
X_1=2 \dot h^2 ((c-2)c+4b-2)-2 (c-2) \left(6 h^2+1\right) \dot h+\left(6 h^2+1\right)^2 \nonumber 
\end{equation}
\begin{eqnarray}
X_2 =&& 6\dot h^2 (c (3 c+10)-4b-6)+\left(6 h^2+1\right)^2 \nonumber \\
&&+6 (c+2)\left(6 h^2+1\right)\dot h, 
\end{eqnarray}
where derivatives are with respect to the dimensionless time in these two expressions.
It is clear that for the RW metric, the Weyl tensor vanishes and the parameter $a$ did not contribute to the reduced action.  Note also that for the particular case $c=-1$ and $b = -5/2$, one has a purely quartic theory since $X_1 =X_2$ whose spherically symmetric solutions were studied in \cite{esin}. Equipped with this, the necessary and sufficient equation to solve is the one coming from 
\begin{equation}
{\frac{ \delta I}{\delta g_{00}}\,\vline}_{RW}  =0,
\end{equation} 
which yields \cite{Weinberg}
\begin{equation} 
 {\cal I} - h \frac{\partial  {\cal I}}{\partial h} + ( -\dot h^2 + 3 h^2)  \frac{\partial  {\cal I}}{\partial \dot h} +h \frac{d }{d \tilde t}\Big (  \frac{\partial  {\cal I}}{\partial \dot h} \Big ) =0.
\label{eom}
\end{equation}
The resulting nonlinear ode is cumbersome and not particularly illuminating to depict here. But, what is important is the following: assuming  a constant Hubble parameter $h(t) = h_0$, the equation reduces to 
\begin{equation}
3 \left(36 h_0^4 \left(3 h_0^2+1\right)-1\right) h_0^2+\lambda_0 =0,
\label{dS}
\end{equation}
which is of course nothing but the vacuum equation (\ref{eff_cosmo}) with  $\lambda = h_0^2/3 $. There is a unique viable solution as long as $\lambda_0 < 11/64$.  For the moment let us assume that this pure de Sitter solution is $h_0$. Then the scale factor grows as  $a(t) = \exp ({ \frac{1}{\sqrt{\gamma}}h_0 t})$. 
This state of affairs cannot go on forever, inflation must end  with a grateful exit after a sufficient  number of $e$-foldings say, $N$, which is around 50-70. For this purpose let us consider the stability of de Sitter phase. Let  $\delta h(t)$ denote a perturbation of the form
\begin{equation}
h(t) = h_0 + \delta h(t) .
\end{equation}
Inserting this in (\ref{eom}) and expanding up to the first order in $\delta h $ and making use of  (\ref{dS}), one arrives at a first order differential equation whose solution is
\begin{equation}
\delta h(t) = c_1 e^{ \frac{1}{\sqrt{\gamma}}\xi h_0 t},
\end{equation}
where 
\begin{equation}
\xi \equiv  \frac{ 12 h_0^2 -1}{ 18h_0^2}.
\end{equation}
Note that  the parameters $c$ and $b$ play a role in the full nonlinear evolution of the Hubble parameter but not its de Sitter value and on the linearization about the de Sitter value. Inflation ends when $\xi >0$ as it shows the instability of the de Sitter solution. But before it ends, the universe expands by an $e$-folding of $N \approx 1/\xi$, which yields
\begin{equation}
N =  \frac{ 18h_0^2}{ 12 h_0^2 -1}.
\end{equation}
Using (\ref{dS}), we can relate the bare dimensionless cosmological constant to the number of $e$-foldings as 
\begin{equation}
\lambda_0= \frac{N (N (11 (N-6) N+108)-54)}{4 (3-2 N)^4}.
\end{equation}
Consider the extreme case of $N \rightarrow \infty$, which yields  $\lambda_0 =\frac{11}{64}$ and $\lambda = \frac{1}{4}$, {it i.e.} the point where the inverse of the effective  Newton's vanish and the solution is pure de Sitter. The discriminant of the quartic equation (\ref{dS}) is  $\Delta = \frac{1}{4} ( 1 + 4\lambda_0)^2 ( -\frac{11}{64} +\lambda_0)$  which is negative for $\lambda_0 < \frac{11}{64}$, hence there are two real and two complex roots and any finite number of desired $e$-foldings can be accommodated given an appropriate $\lambda_0$. Figure 1 shows the relation.
\begin{figure}[h]
\includegraphics[width=7 cm]{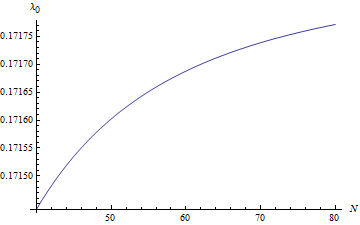}\caption{Bare cosmological parameter versus number of $e$-foldings }
\end{figure}
There is one final piece of puzzle we have to settle: number of $e$-foldings when written in terms of $\lambda$ is simply
\begin{equation}
N = \frac{6 \lambda}{4 \lambda -1},
\end{equation}
which requires  $\lambda > \frac{1}{4}$.  Demanding $G_0 >0$, which is the Newton's constant at low energies right after the inflation, one must have the effective Newton's constant $G_N < 0 $ at high energies. The fact that the Newton's constant run with curvature an becomes negative as the (scalar) curvature is   $ R \ge  4 \Lambda =\frac{\lambda} {\gamma}$, gives  the required repulsive 
era for gravity. During the inflation, the curvature is reduced and the Newton's constant start to grow which provides the needed "friction" mechanism or stopping the inflationary period. 
Once the curvature is diluted and the universe becomes flat, one has the usual  positive Newton's constant $G_0$ which can be taken numerically as today's  value.
To give some numerical values,  let us  demand $N= 70$ $e$-foldings, then, $G_N \approx -20\,  G_0 $.  If the Inflation occurred at a high energy scale $E_I = 2 \times 10^{16}$ GeV,  then our dimensionful parameter $\gamma$ is around this value, basically the relation is  $\gamma  = \frac{ \lambda}{3 H^2_{inf}}$, which is numerically a very small value making sure that higher derivative terms do not dominate when the curvature is small, after the inflation.

\section{Conclusions and Discussions}
  
We studied the  inflationary phase of a pure higher derivative  gravity theory that has the same particle spectrum as general relativity, namely only a massless spin-2 graviton. The effective Newton's constant is a function of the curvature and in the maximally symmetric case of the Robertson-Walker spacetime, a function of the effective cosmological parameter, a dimensionless number, we we called $\lambda$ that decreases as the universe expands.  When the curvature is high in the pre-inflation moments of the universe, higher derivative terms dominate and the theory is repulsive, but as the universe expands, the curvature becomes flatter, as $\lambda$ decreases and the effective Newton's constant start to grow and becomes positive after passing through the fixed point given by  $\lambda = 1/4$.   This allows a rather interesting freedom for the theory:  a quasi-de Sitter solution with any number of $e$-foldings is possible depending on the value of the bare cosmological constant. The theory is a close cousin of Einstein's gravity in many aspects but, unlike cosmological Einstein's gravity, it finds a way out of never-ending inflation due the above mentioned behavior of the effective Newton's constant.

What we have presented here is a suggestion that pure gravity with higher derivative terms, that has the same perturbative spectrum as  Einstein's theory and no other additional fields, can  can have a period of inflation and an exit from inflation after sufficient number of $e$-foldings. [See \cite{Pani} for another point of view where gravity is that of Einstein's theory coupled to  non-dynamical (auxiliary) fields that necessarily involve  higher derivative.] for The theory, presented here is to be considered as an effective theory with higher curvature terms playing a major role during inflation and it can be expanded  by adding more powers inside the determinant while keeping the particle spectrum intact. This will also change the behavior of the effective Newton's constant. After the higher derivative terms play their role, and the curvature is diluted,  the ensuing theory is the  Einstein's gravity coupled to sources, such as the standard model fields or the perfect fluid. For the theory to be a viable one, one must work out the density perturbations during inflation: for this purpose there are various possibilities of minimal or non-minimal coupling as our theory is a  determinantal form.

\section{Acknowledgment}

 This work was supported by the TUB\.{I}TAK grant 113F155. I would like to thank  M. Gurses, A. Karasu, P. Olver, A. Erkip and O. Kisisel for  useful discussions  on nonlinear ode and A. Kaya, E. O. Kahya and T.C. Sisman on inflation and related matters.

\end{document}